\documentclass[twocolumn,showpacs,prl,aps]{revtex4}
\usepackage{graphicx}
\usepackage{dcolumn}
\usepackage{bm}
\usepackage{latexsym}
\usepackage{amssymb}
\usepackage{amsmath}
\usepackage{graphics}
\usepackage{subfigure}

\begin{document}

\title{Causality relations for materials with strong artificial optical chirality}

\author{M.V. Gorkunov$^1$, V.E. Dmitrienko$^1$, A.A. Ezhov$^{1,2,3}$, V.V. Artemov$^1$, O.Y. Rogov$^1$}
\affiliation{$^1$A.V. Shubnikov Institute of Crystallography, Russian Academy of Sciences, 119333 Moscow, Russia\\
$^2$M.V. Lomonosov Moscow State University, 119991 Moscow, Russia\\
$^3$A.V.Topchiev Institute of Petrochemical Synthesis, Russian Academy of Science, 119991 Moscow, Russia}

\begin{abstract}
We demonstrate that the fundamental causality principle  being applied to strongly chiral artificial materials yields the generalized Kramers-Kronig relations for the observables -- circular dichroism and optical activity. The relations include the Blaschke terms determined by material-specific features - the zeros of transmission amplitude on the complex frequency plane. By the example of subwavelength arrays of chiral holes in silver films we show that the causality relations can be used not only for a precise verification of experimental data but also for resolving the positions of material anomalies and resonances and quantifying the degree of their chiral splitting.
\end{abstract}
\maketitle

The absence of mirror symmetry being a common attribute of numerous natural objects and materials, especially those of biological origin, typically gives rise to very moderate optically observable consequences. Recent progress in the field of artificial chiral electromagnetic materials has shown that one can achieve anomalously high and even extreme values of optical activity (OA) and circular dichroism (OD) \cite{Rogacheva, Decker, Singh, Gibbs, Dietrich, Dave, we}. Remarkably, it also appears to be possible to gain additional functional properties and create high chirality that is 
strongly nonlinear \cite{ren}, invertible by irradiation \cite{zhang}, or appears via spontaneous symmetry breaking \cite{liu}. Such features, being unattainable with natural materials, are highly advantageous for the applications that range potentially from electromagnetic signal manipulation \cite{beamsplitter} to nanoscale chirality diagnostics \cite{Hendry}.

For an artificial structure to acquire a substantial electromagnetic chirality, it has to possess a pronounced structural chirality with a possibly high intrinsic electromagnetic contrast, i.e., it has to include the constituents with sufficiently different electromagnetic response. The structure period has to be subwavelength in order to provide the effective homogeneity of the material. Chiral metamaterials -- subwavelength metal-dielectric structures and arrays with broken mirror symmetry -- have proved to be a very fruitful concept \cite{twist,Valev}.

While the fabrication of illustrative metamaterial samples operational in the radio and microwave ranges does not require the use of sophisticated techniques, creating the structures with micron and submicron periodicity for the infrared and visible ranges still remains challenging and different types of chiral metamaterials obtained by different approaches have been reported: micron-scale double layer structures operational at the wavelengths of a few microns \cite{Decker}, nanoscale dielectric helical templates decorated with plasmonic nanoparticles \cite{Singh}, metallic helices \cite{Gibbs} exhibiting noticeable CD in the visible, precisely elevated  starfish-shaped metal particles \cite{Dietrich} and chiral holes with extreme optical CD and OA \cite{we}.

Independently of the scale and type of chiral medium, the CD and OA are always the key characteristics either being observables for the chirality diagnostics or defining the main functional properties in the prospective applications. It has been recognized for decades that the general principle of causality in the form of appropriate Kramers-Kronig (KK) integral relations for the difference of refractive indices of left and right circularly polarized waves can provide a valuable opportunity to relate CD and OA of natural materials with molecular-scaled inner structure and weak optical chirality (see e.g.  Chapter 21 in Ref. \onlinecite{KingVol2} and refs. therein). In artificial materials, however, the much larger inner scales prevent from introducing effective macroscopic parameters, while on the other hand, the reported spectral behavior of CD and OA is often very far from trivial and seemingly contradicts to the traditional rules of KK-relations. According to them, a resonant peak of one characteristic should be accompanied by an antiresonant kink of its counterpart. While in some artificial media this holds true (see e.g. Ref. \onlinecite{Decker}), in the others the situation is different and a broadband OA may appear with negligibly small CD \cite{Dave} or both OA and CD can peak together up to their extreme values at very close wavelengths \cite{we}.

In this Letter we show that the causality allows introducing an appropriate form of the KK integral relations for the OA and CD. Being strictly correct from the mathematical point of view and based solely on the fundamental principle of causality, the relations can be widely used as a solid reference point. In addition, the correct form of the relations includes the so-called Blaschke terms that are determined by the inner resonances and anomalies of the chiral material. This provides a unique opportunity of extracting valuable quantitative information on important intrinsic material features by means of conventional spectropolarimetry.
Using as example the spectropolarimetry data for the arrays of nanosize chiral holes in metal films we demonstrate the latter  possibility and obtain explicitly the complex eigenfrequencies of the chirally split leaky waves supported by the arrays.

Mathematically, the KK-relations connect the real and imaginary parts of a function of complex variable that is known to be analytical in the upper half-plane of the variable. In physics, a direct and simple deduction to the causality principle allows applying the KK-relations to calculate the frequency dependence of the real parts from the known imaginary parts (or vice versa) of refractive index, permittivity or susceptibility \cite{Landau}.
However, these well-known forms of the KK integrals are not universal and are to be replaced by more general relations if the analytical properties of the response functions are more complicated.

In particular, it often useful to consider the logarithm of the reflection or transmission amplitudes as  response functions. If the amplitudes turn to zero (staying analytical) at certain complex frequencies $\omega_i$ in the upper-half complex frequency plane, the KK-relations for the logarithms are to be modified accordingly by introducing the so-called Blaschke term.
Appearance of such situations in various physical problems was first recognized by van Kampen \cite{Kampen53} and then analyzed in detail by Toll \cite{Toll56}. The Blaschke term changes drastically the phase of the reflection and transmission coefficients hence being of key importance in the phase retrieval problems \cite{Nussenzveig}.

For what follows it is necessary to apply the causality principle to the transmission problem and we consider  the logarithm of the transmission amplitude $t(\omega)= |t(\omega)|\exp[i\Psi(\omega)]=
\exp[\ln|t(\omega)|+i\Psi(\omega)]$. Then, if $\ln|t(\omega)|$ is analytical everywhere in the upper half-plane of complex $\omega$, the
KK-relations read:
\begin{equation}\label{KK1}
\ln|t(\omega)|=-\frac{1}{\pi}\ \text{P}\int_{-\infty}^\infty\
\frac{\Psi(\omega')\ d\omega'}{\omega'-\omega},
\end{equation}
\begin{equation}\label{KK2}
\Psi(\omega)=\frac{1}{\pi}\ \text{P}\int_{-\infty}^\infty\
\frac{\ln|t(\omega')| d\omega'}{\omega'-\omega},
\end{equation}
where P stands for the principal value of the integrals.

If $t(\omega)$ has one or more zero points in
the upper-half plane of the complex frequency, one can  consider an auxiliary function $\hat t$ such that  $t(\omega)=B(\omega)\hat t(\omega)$, where the Blaschke multiplier $B(\omega)$ contains explicitly all $n\ge 1$ zeros $\omega_i$ of
$t(\omega)$ (multiple roots considered as different roots):
\begin{equation}\label{B}
B(\omega)=\prod_{i=1}^n \frac{\omega-\omega_i}{\omega-\omega_i^*}
\end{equation}
where the star means complex conjugate. Obviously $|B(\omega)|=1$ for real $\omega$ and the multiplier changes only the phase of the transmission amplitude. The
analytical function $\hat t(\omega)$ has no zeros and for its logarithm one can write the KK-relations. As a result, the KK-relations \eqref{KK1} and \eqref{KK2} for the physically meaningful $|t|$ and $\Psi$ are to be adjusted by the substitution
\begin{eqnarray}\label{Psi1}
\Psi(\omega)\rightarrow\Psi(\omega) + \sum_{i=1}^n
\arg\left(\frac{\omega-\omega_i}{\omega-\omega_i^*}\right).
\end{eqnarray}

For a chiral system one applies the above routine to the transmission amplitudes $t_R$ and $t_L$ of the right and left circularly polarized radiation respectively. Then the direct relations between the observable CD defined conventionally as $D=(|t_{R}|^2-|t_{L}|^2)/(|t_{R}|^2+|t_{L}|^2)$ and the OA rotation angle $\Phi=[\arg(t_{R})-\arg(t_{L})]/2$ can be established:
\begin{equation}\label{KKch1}
\ln\frac{1+D(\omega)}{1-D(\omega)}=-\frac{8\omega^2}{\pi}\ \text{P}\int_0^\infty\ \frac{\hat\Phi(\omega')\ d\omega'}{\omega'(\omega'^2-\omega^2)}
\end{equation}
\begin{equation}\label{KKch2}
\hat\Phi(\omega)=\frac{\omega}{2\pi}\ \text{P}\int_0^\infty\ln\frac{1+D(\omega')}{1-D(\omega')}\ \frac{d\omega'}{\omega'^2-\omega^2}
\end{equation}
where the angle entering the KK-relations reads
\begin{multline}\label{Bch}
\hat\Phi(\omega)=\Phi(\omega)+\frac{1}{2}\sum_{i=1}^{n_R}
\arg\left(\frac{\omega-\omega_{Ri}}{\omega-\omega^{*}_{Ri}}\right)-\\
\frac{1}{2}\sum_{i=1}^{n_L}
\arg\left(\frac{\omega-\omega_{Li}}{\omega-\omega^{*}_{Li}}\right).
\end{multline}
Here the summations over the zeros $\omega_{Ri}$ and $\omega_{Li}$ of the amplitudes $t_R$ and $t_L$ correspondingly is performed.

\begin{figure}
\centering
\includegraphics[width=8.4cm]{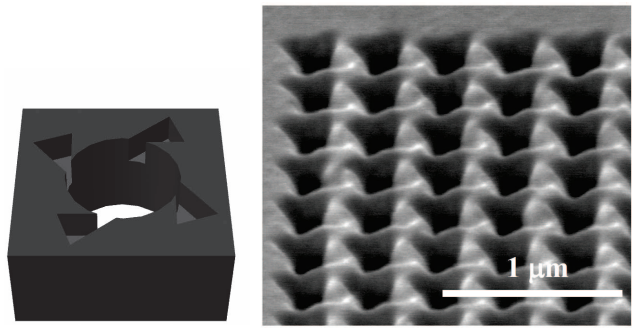} \caption{3D-model of a chiral hole as implemented into the focused ion beam milling digital template (left) and SEM image of a fabricated array A1 tilted by  52$^{\rm o}$ (right).}\label{fig1}
\end{figure}

To illustrate the application of these very general relations, we consider the experimental data on strong OA and CD exhibited by subwavelength 4-fold arrays of chiral holes in freely suspended silver foils. Technically, the experimental methods of fabrication and optical characterization used have been very similar to those reported recently in Ref.~\onlinecite{we}.
Two sample arrays discussed below are of the same type, have equal lateral dimensions and were milled using FEI Helios DualBeam microscope in the foils of different thickness: 270 nm (array A1) and 380 nm (array A2).
Being fabricated with single-pass focused ion beam (FIB) milling according to the digital template shown schematically in Fig.~\ref{fig1} on the left, the arrays possess the in-plane fourth order rotational symmetry. A fragment of the fabricated array A1 is shown in Fig.~\ref{fig1} on the right where the   difference from the template due to ion beam defocusing and diversion can be seen as well. The periods of both array square lattices were set to 375 nm to avoid diffraction in the visible, and the inner hole diameter was 187 nm. The symmetry breaking responsible for the structural chirality (the absence of  mirror planes) was granted by the offset of the triangles patterned on one array interface.

\begin{figure}
\includegraphics[width=8.5 cm]{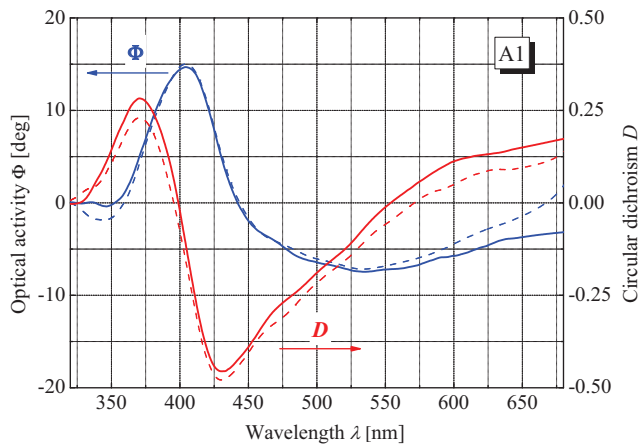}
\caption{Measured CD and OA of the array A1 (solid lines) compared to  CD and OA calculated according to Eqs.~\eqref{KKch1} and \eqref{KKch2} without Blaschke terms (dashed).}\label{fig2}
\end{figure}

\begin{figure}
\includegraphics[width=8.5 cm]{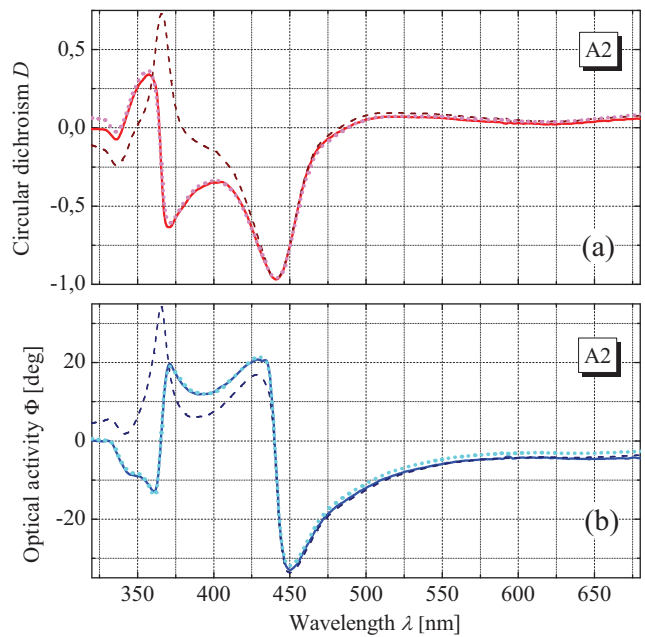}
\caption{Measured CD (a) and OA (b) of the array A2 (solid lines) compared to CD and OA calculated according to Eqs.~\eqref{KKch1} and \eqref{KKch2} without (dashed) and with (dotted) Blaschke terms being taken into account.}\label{fig3}
\end{figure}

The microspectropolarimetry of light transmitted through the samples was carried out with a spectroscopic Uvisel 2 (Horiba Jobin-Yvon) ellipsometer as described in Ref.~\onlinecite{we}. The CD and OA data obtained for the samples A1 and A2 are shown by solid lines in Figs.~\ref{fig2} and \ref{fig3} respectively. The optical chirality of both arrays is notably strong as the OA reaches several tens of degrees and the CD peakes down to the value of $-0.5$ in the thinner array A1 and even reaches the extremal value of $-1$ in the thicker array A2. Remarkably, although the complex spectral behavior of the strong optical chirality seems to have much in common with the recently reported extreme optical chirality of 4-start screw thread chiral holes \cite{we}, there exists a qualitative difference seen vividly in Fig.~\ref{fig2}, where a peak of OA is accompanied by an antiresonance of CD. This situation is inverse compared to the data for the threaded holes \cite{we} and also appears to be rather unusual in general as in natural chiral materials thin peaks of handedness-selective absorption give rise to antiresonant OA.

Next, we have calculated the integrals in Eqs.~\eqref{KKch1} and ~\eqref{KKch2} numerically for the experimental data.
The results of this integration obtained without accounting for possible Blaschke phase are presented as dashed lines in Figs.~\ref{fig2} and \ref{fig3} and demonstrate clearly  the real complexity of the application of casuality principle to artificial chirality. The OA and CD spectra for the thinner A1 array satisfy the simple KK-relations nicely, as the calculated and measured values practically coincide in the broad visible range (small deviations closer to the borders are related to the finite spectral range of experimental data).
At the same time, the data obtained for the thicker array A2 show a dramatic discrepancy. In fact, even without doing calculations one can see that the behavior of CD and OA for the array A2 is counterintuitive: Both quantities experience a pronounced antiresonance around the 370 nm wavelength. According to the conventional KK-relations, an antiresonance of one quantity has to be accompanied by a resonance of its counterpart and the dashed lines in Fig.~\ref{fig3} behave exactly in this manner being in total disagreement with the experiment.

\begin{figure}
\includegraphics[width=8.5 cm]{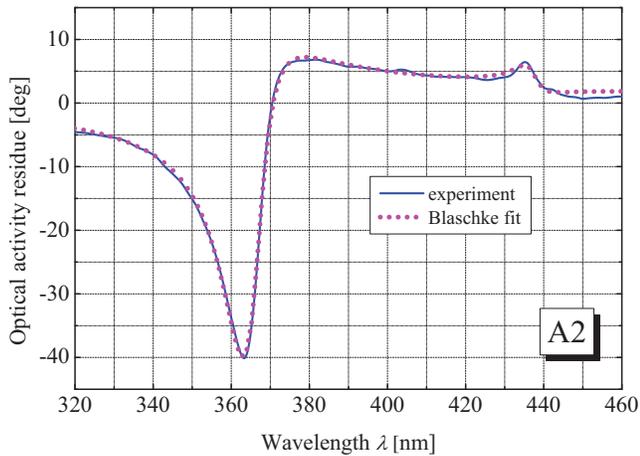}
\caption{Difference of the measured OA $\Phi$ and $\hat\Phi$ calculated from the measured CD according to Eq.~\eqref{KKch2} for the array A2 (solid) fitted by the Blaschke phase \eqref{Bchlamb} (dotted) with the parameters $\lambda_{R1}=363.6-7.3i$ nm, $\lambda_{L1}=366.6-4.0i$ nm, $\lambda_{R2}=437.1-3.2i$ nm, $\lambda_{L2}=436.9-3.3i$ nm.}\label{fig4}
\end{figure}

To clarify the origin of such situation it is useful to consider the difference between the observed $\Phi$ and $\hat\Phi$ calculated from the measured CD according to Eq.~\eqref{KKch2}. As shown in Fig.~\ref{fig4}, this residue has a very particular form being obviously a sum of narrow peaks around specific wavelengths. Notably, it can be fitted precisely with the Blashke phase \eqref{Bch} that as a function of wavelength reads
\begin{multline}\label{Bchlamb}
\Phi(\lambda)-\hat\Phi(\lambda)=\sum_{i=1}^{n_R}
\arg\left(\frac{\lambda-\lambda_{Ri}}{\lambda_{Ri}}\right)-
\sum_{i=1}^{n_L}
\arg\left(\frac{\lambda-\lambda_{Li}}{\lambda_{Li}}\right),
\end{multline}
where $\lambda_{R,Li}=2\pi c/\omega_{R,Li}$.
As seen in Fig.~\ref{fig4}, the Blaschke phase with the contributions from two pairs of the transmission zero points on the complex plane appears to be sufficient for the fitting.
Implementing thus resolved zero points into the generalized KK-relations (\ref{KKch1}--\ref{Bch}) allows calculating the OA from the CD and vice-versa that nicely coincide with the experiment (see dotted lines in Fig.\ref{fig3}).

The obtained quantitative information on special complex frequencies (wavelengths) of the transmission zeros is in fact a very valuable knowledge that is hard to be extracted otherwise. The strongest contribution to the Blaschke phase is provided by the first pair of zero points, $\lambda_{R1}$ and $\lambda_{L1}$, which real parts are very close to the array period (375nm$\pm 5\%$  according to the SEM images). Since decades it has been recognized that such special points are responsible for the Wood anomaly adjacent to the diffraction Rayleigh anomaly and determined by the existence of leaky surface waves supported by metallic arrays and gratings \cite{Hessel}. Here we provide eventually a very convenient approach to extract the exact location of such points from the experiment. Moreover, we reveal also that the structural chirality causes a noticeable chiral splitting of the guided modes spectrum and substantially affects the guided modes quality factor  (as the mode $\lambda_{R1}$ has almost a twice lower quality factor than the mode $\lambda_{L1}$).

Interestingly, we have also resolved another pair of special points, $\lambda_{R2}$ and $\lambda_{L2}$, that are located much closer to each other around the wavelength of 437 nm. The almost negligible chiral splitting of the points makes their contribution to the Blaschke phase noticeably weaker and reveals a much lower sensitivity of the underlying resonances to the overall structure chirality. As also the wavelength of 437 nm differs from the characteristic array sizes (thickness and period) we presume that these resonances are well localized plasmonic resonances of the complexly shaped metal structure.

For other artificial chiral structures, the formulated generalized causality relations can clarify various important questions. One such question is related to the possibility of strong electromagnetic chirality in perfectly conducting chiral structures of high (4-fold) rotational symmetry. It has been shown very generally that the CD can arise in such structures only accompanied by losses \cite{we, Kaschke} and thus a lossless structure possesses an almost negligible CD (as observed  e.g. in Ref. \onlinecite{Dave}). On the other hand, the simple KK-relation \eqref{KKch2} suggests that there has to be also no OA in such case, which is apparently against the observations. It is the Blaschke phase that can resolve this paradox as the total OA of perfectly conducting chiral structure has to exhibit the simple dispersion \eqref{Bchlamb} with appropriate set of transmission zero points.

Finally, our findings may also have an important application to soft chiral materials such as cholesteric and blue-phase liquid crystals and polymers. It has been shown recently \cite{Dolganov} that in the simple planar geometry a cholesteric layer produces CD and OA that obey quite precisely the simple KK-relations \eqref{KKch1} and \eqref{KKch2}. In other geometries, when the Bragg vector is parallel to the entrance surface, one can expect appearance of the Blaschke term in the phase of transmitted wave. In this case, the transmission zeroes should be determined by the topological charge of the diffraction band (defined similarly to that of the x-ray diffraction bands \cite{Dmitrienko}).

In conclusion, we have shown that a strict application of the causality principle to strongly chiral artificial materials is possible in the form of generalized KK-relations. Being formulated for the main observables -- CD and OA, the relations include the Blaschke terms determined by the transmission zero points on the complex frequency plane. Using as example the data for strongly chiral optical response of subwavelength arrays of chiral holes in silver films we have demonstrated that the KK-relations can be used for a precise verification of experimental data and, more significantly, for resolving the discrete spectrum of material-specific anomalies and resonances.

We are grateful to S.G. Yudin for silver sputtering, to  A.L. Vasiliev, Shared Research Centers of IC RAS and MSU supported by the Ministry of Education and Science of the Russian Federation for the equipment provided. The work was supported by the Russian Science Foundation (project 14-12-00416).


\end{document}